\documentclass[conference]{IEEEtran}
\IEEEoverridecommandlockouts
\usepackage{cite}
\usepackage{amsmath,amssymb,amsfonts}
\usepackage{algorithmic}
\usepackage{graphicx}
\usepackage{textcomp}
\usepackage{xcolor}
\usepackage{comment}

\usepackage{siunitx}
\usepackage{hyperref}
\usepackage[nolist]{acronym}
\usepackage[flushleft]{threeparttable}
\usepackage{placeins}
\usepackage{svg}                                   
\usepackage[percent]{overpic}

\usepackage{tabularx}
\usepackage{caption}
\usepackage{booktabs}
\usepackage[export]{adjustbox}

\usepackage[normalem]{ulem}

\def\BibTeX{{\rm B\kern-.05em{\sc i\kern-.025em b}\kern-.08em
    T\kern-.1667em\lower.7ex\hbox{E}\kern-.125emX}}
\begin{document}
\bstctlcite{IEEEexample:BSTcontrol}

\newcommand{\todo}[1]{\textcolor{red}{#1}}                          
\newcommand{\done}[1]{\textcolor{Green}{#1}}                        
\DeclareSIUnit{\dBm}{\text{dBm}}                                    
\newcommand{\RNum}[1]{\uppercase\expandafter{\romannumeral #1\relax}}  

\title{Towards a Novel Ultrasound System Based on Low-Frequency Feature Extraction From a Fully-Printed Flexible Transducer}
\author{
\IEEEauthorblockN{
Marco Giordano\IEEEauthorrefmark{1}, Kirill Keller\IEEEauthorrefmark{2}, Francesco Greco\IEEEauthorrefmark{2}\IEEEauthorrefmark{4}, Luca Benini\IEEEauthorrefmark{1}\IEEEauthorrefmark{3}, Michele Magno\IEEEauthorrefmark{1}, Christoph Leitner\IEEEauthorrefmark{1}
}
\IEEEauthorblockA{\IEEEauthorrefmark{1} Center for Project Based Learning/Integrated Systems Laboratory, ETH Z{\"u}rich, Z{\"u}rich, Switzerland}
\IEEEauthorblockA{\IEEEauthorrefmark{2} Institute of Solid State Physics, Graz University of Technology, Graz, Austria}
\IEEEauthorblockA{\IEEEauthorrefmark{3} DEI, University of Bologna, Bologna, Italy}
\IEEEauthorblockA{\IEEEauthorrefmark{4} BioRobotics Institute, Sant'Anna School of Advanced Studies, Pisa, Italy}
\vspace{-1.cm}
\thanks{Corresponding author: \{christoph.leitner\}@iis.ee.ethz.ch}
}

\maketitle



\begin{abstract}
Ultrasound is a key technology in healthcare, and it is being explored for non-invasive, wearable, continuous monitoring of vital signs. However, its widespread adoption in this scenario is still hindered by the size, complexity, and power consumption of current devices. Moreover, such an application demands adaptability to human anatomy, which is hard to achieve with current transducer technology. This paper presents a novel ultrasound system prototype based on a fully printed, lead-free, and flexible polymer ultrasound transducer, whose bending radius promises good adaptability to the human anatomy. 
Our application scenario focuses on continuous blood flow monitoring.

We implemented a hardware envelope filter to efficiently transpose high-frequency ultrasound signals to a lower-frequency spectrum. This reduces computational and power demands with little to no degradation in the task proposed for this work. We validated our method on a setup that mimics human blood flow by using a flow phantom and a peristaltic pump simulating 3 different heartbeat rhythms: $\mathbf{60}$, $\mathbf{90}$ and $\mathbf{120}$ beats per minute.  Our ultrasound setup reconstructs peristaltic pump frequencies with errors of less than $\mathbf{0.05}$ Hz ($\mathbf{3}$ bpm) from the set pump frequency, both for the raw echo and the enveloped echo. The analog pre-processing showed a promising reduction of signal bandwidth of more than 6x: pulse-echo signals of transducers excited at $\mathbf{12.5}$ MHz were reduced to about $\mathbf{2}$ MHz. Thus, allowing consumer MCUs to acquire and elaborate signals within mW-power range in an inexpensive fashion. 
\end{abstract}

\vspace{10pt}


\vspace{-0.5cm}
\begin{figure*}[ht]
  \centering
\includegraphics[width=0.8\textwidth,trim=0 0 0 0,clip]{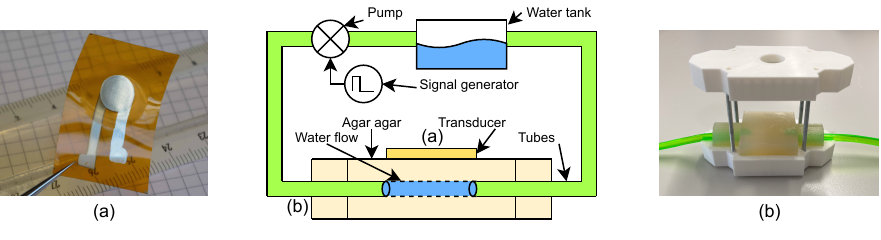}
\caption{Schematic of the system setup. (a) shows the flexible transducer. (b) shows the phantom inside the 3D-printed mold.}
    \label{fig:system_setup}
  \vspace{-0.4cm}
\end{figure*}

\section{Introduction}

\renewcommand{\arraystretch}{1.2}

\makeatletter
\def\ps@IEEEtitlepagestyle{
  \def\@oddfoot{\mycopyrightnotice}
  \def\@evenfoot{}
}
\def\mycopyrightnotice{
  \begin{minipage}{\textwidth}
  \centering \scriptsize
  Copyright~\copyright~2023 IEEE. Personal use of this material is permitted. Permission from IEEE must be obtained for all other uses, in any current or future media, including\\reprinting/republishing this material for advertising or promotional purposes, creating new collective works, for resale or redistribution to servers or lists, or reuse of any copyrighted component of this work in other works.
  \end{minipage}
}
\makeatother

Ultrasound technology, long utilized as a diagnostic tool, is witnessing a notable expansion into wearable applications, fueled by ongoing advancements in material sciences \cite{j:lin2023}, system design \cite{c:frey2022} and signal processing \cite{c:brausch2019}. As an echographic imaging technique, it carries the unique ability to penetrate soft tissues and capture real-time images in a radiation-free manner, enabling doctors to assess health conditions without resorting to invasive procedures. However, the potential benefits of ultrasound extend beyond traditional healthcare applications. Recent works show possible applications of ultrasound technology in the realm of wearable monitoring of hemodynamics \cite{j:hu2023} and organs \cite{j:wang2022, j:lin2023}, in prosthetic controls \cite{j:yin2022} as well as in human-machine interfaces (HMI) \cite{ax:grandi2023}. Exploiting novel materials, compact hardware designs, and advanced signal processing techniques, ultrasound technology is now being integrated into smart patches, with the ultimate goal of inconspicuously obtaining vital data from tissues and internal organs \cite{c:vostrikov2023}, representing an innovative, non-invasive approach to patient monitoring and personalized healthcare.

Recently, flexible ultrasonic transducers \cite{j:hu2023} and even bendable 2D arrays \cite{j:wang2022} based on lead zirconate titanate (PZT) have been proposed. Lin et al. \cite{j:lin2023} demonstrated the first fully wearable ultrasound system tailored for transmitting and receiving signals from a thin and flexible PZT ultrasound transducer. PZT has excellent piezoelectric properties, but their fabrication is complex and transducers contain harmful lead. 

Piezoelectric polymers such as P(VDF-TrFE) provide a compelling alternative to materials like PZT. Not only is P(VDF-TrFE) lead-free, but it is also significantly easier and more cost-effective to manufacture, thanks to its compatibility with printing methods. These characteristics make P(VDF-TrFE) a promising candidate for its use in wearable ultrasonic devices\cite{c:leitner2022}.
To date, P(VDF-TrFE)-based transducers have been used primarily for high-frequency applications, but they are also finding their way into medical application scenarios where flexibility and conformability are key. The inherent advantages of P(VDF-TrFE) including its high sensitivity and wide bandwidth, can make it a key component in transferring ultrasound instruments from traditional benchtop or handheld hardware form factors into wearable devices\cite{pvdf_advantages}.

One of the challenges with screen-printed P(VDF-TrFE) is the feasible (often thin) layer thicknesses \cite{j:wagle2013,c:leitner2022}. This results in high resonant frequencies \cite{j:brown2000} that might not be suitable for medical ultrasound and pose a challenge for signal acquisition and processing hardware. From a fabrication viewpoint, care must be taken in the material selection and thickness choice of auxiliary structures (e.g. substrate, electrodes,...) as they all acoustically modulate the transducer bandwidth \cite{c:leitner2023}. Moreover, the use of such transducers in wearable scenarios requires custom-made ultrasonic acquisition system designs which can bypass power-hungry electronics that are usually required for high-frequency signal acquisition \cite{c:frey2022}.

In this context, and to address the above challenges, we present a design concept for an ultrasound acquisition platform. We developed and evaluated an envelope filter strategy to extract features from the ultrasound signal, transposing high-frequency ultrasound signals to a low-frequency spectrum, thus reducing computational and power demands without compromising vital diagnostic data. We incorporate a flexible high-frequency ultrasound transducer based on P(VDF-TrFE). Moreover, we employ an integrated pulser in a compact form factor that ensures effective ultrasound signal generation while enhancing portability and energy efficiency. We demonstrate the application of our setup on hemodynamic measurements in a phantom.
Our ultrasound setup manages to reconstruct peristaltic pump frequencies with errors of less
than \SI{0.05}{\Hz} (\SI{3}{bpm}) from the set pump frequency of 1-\SI{2}{\hertz} (60-\SI{120}{bpm}), both for the raw echo and the enveloped echo. 
Our system prototype allows for easy fabrication and has the potential for miniaturization, other than opening doors towards low-power ultrasound signal conditioning and onboard processing for wearable devices.

In conclusion, this paper demonstrates that a low-power front-end for ultrasound can be used to lower system requirements for high-frequency fully printed ultrasound transducers. The technology was tested on a hemodynamic phantom, showing great potential for continuous blood flow monitoring.

\begin{figure*}[t]
\begin{minipage}{0.58\textwidth}
\centering
\includegraphics[width=0.9\textwidth]{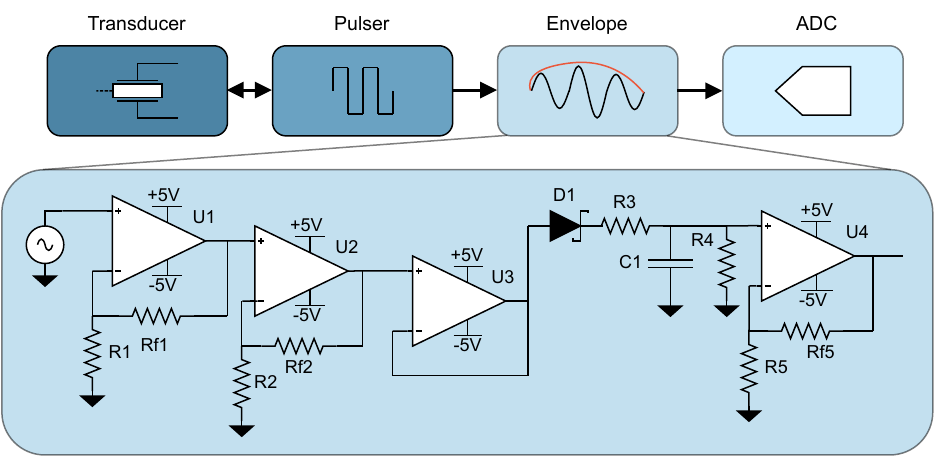}
\caption{Schematic of the signal acquisition pipeline and detail on the envelope detection circuit.}
\label{fig:circuit}
\end{minipage}\hfill
\begin{minipage}[t]{0.39\textwidth}
\vspace{-3cm}
\begin{minipage}[t]{\textwidth}
\centering
\begin{tabular}{ccc}
Ground truth & Raw [Hz] & Envelope [Hz] \\
\midrule
60bpm - 1Hz                       & 0.95 & 1.01       \\
90bpm - 1.5Hz                     & 1.45 & 1.45       \\
120bpm - 2Hz                      & 2.03 & 1.98     
\end{tabular}
\captionof{table}{Peak detection over the accumulated FFT per phantom frequency.}
\label{tab:results_statistics}
\end{minipage}
\begin{minipage}[t]{\textwidth}
\vspace{0.5cm}
\centering
\begin{tabular}{ccc}
& Pulser & Envelope \\
\midrule
Power & \SI{1.9}{\milli\watt} & \SI{0.14}{\milli\watt} 
\end{tabular}
\captionof{table}{Estimated system power.}
\label{tab:pow_estimation}
\end{minipage}
\end{minipage}
\vspace{-0.5cm}
\end{figure*}



\section{Materials and Methods}\label{sec:methods}

An overview of the setup is summarised in \autoref{fig:system_setup}. The flexible transducer, seen in detail in \autoref{fig:system_setup}(a), is placed on top of the phantom, \autoref{fig:system_setup}(b), while water is pumped through the phantom with the peristaltic pump. In the following subsections, the different components of the setup as well as the experimental measurement setup will be introduced and discussed.

\subsection{Flexible Polymer Transducer Characterization}
For this study, we employed a P(VDF-TrFE) ultrasonic transducer printed on a flexible substrate (\autoref{fig:system_setup}(a). As the printing substrate we used Kapton HN (DuPont) with \SI{50}{\um} thickness. P(VDF-TrFE) transducers were screen printed using FC20 ink (Arkema). Electrodes were inkjet printed using reactive silver ink EI-502 (Electroninks). The entire fabrication process has already been described in detail in Leitner et al. \cite{c:leitner2022}. Our transducers showed the resonance frequency at \SI{15.1}{\MHz}, they were circular and had an active surface area of \SI{56.75}{\mm^2}.

\subsection{Transmit and Receive Circuit}
\subsubsection{Pulser}
Targeting an embedded implementation, we aimed for a low-power pulser with a small device footprint, which could deliver the necessary power for our transducer. The choice was the STHVUP32 (STMicroelectronics), a novel pulser technology that provides the ability to design custom waveforms on up to 5 different voltage levels for 32 different channels, coordinated by an internal beamforming engine.

\subsubsection{Envelope detection}
Ultrasound applications are characterized by fairly high-frequency signals, usually in the MHz range. To cope with such high data rates in conventional medical devices, dedicated ADCs are used to acquire raw signals which are then sent to FPGAs for further processing\cite{ultrasound_fpga}. Therefore, the components required are expensive and do not target low-power operation in the mW-range. Since we expected the operating frequencies of our printed P(VDF-TrFE) transducers to be in the range of 10-15 MHz one core idea of this work was to pre-process the ultrasound signals in the analog domain before the digital conversion. We experimented with extracting the envelope of the receiving ultrasound signal in hardware, validating the approach in a frequency extraction task. The proposed approach allows to drastically reduce the data rates required for the acquisition signals from the P(VDF-TrFE) transducers. Thus, minimizing the power consumption and system requirements for the signal processing, at a mW-range power expense.

\autoref{fig:circuit} shows the envelope detection circuit implemented in this work. The operational amplifier (opamp) chosen is the ADA4807 (Analog Devices) due to its analog characteristics compatible with the ultrasound signal and its low-power operation. The first two opamps, U1 and U2, are non-inverting amplifiers to amplify and shift the echo signal. Moreover, given the limited gain-bandwidth product of the opamps, they also filter (by having a smaller amplification factor) the high-frequency noise caused by the pulser. The following opamp, U3, serves to decouple the feedback networks of the first two amplifiers from the envelope detection circuit. The envelope detection starts from D1, which feeds a rectified signal into R3 that in turn charges the capacitor C1, which is then discharged by R4. The resulting signal is once again amplified by U4, which decouples the envelope detection circuit from the signal acquisition pipeline.

\subsection{Flow Phantom}
\label{sec:phantom}
To verify our experimental setup within a regulated setting, we engineered a phantom made from agar-agar that simulates a blood vessel \cite{EARLE201618}. 
Our phantom was produced using a 3D-printed mold that encompasses a tube along its primary axis. Following the curing process of the material, this tube is extractable, allowing for the attachment of an additional pair of tubes at the endpoints. This process effectively forms a conduit within the agar-agar, which is subsequently filled with water. The section of the phantom, as well as a picture of the mold, can be seen in \autoref{fig:system_setup}.
To emulate the rhythmic beating of the heart, we employed a peristaltic pump. The pump is operated under a duty cycle to replicate a single heartbeat for each period, thereby mirroring the human's heart beating. Our objective is to reconstruct the pump frequency, which emulates the heart rate, from the ultrasound data collected with our system. 

\subsection{Flow Measurements}

In this section, the experimental setup and the signal processing used to retrieve the peristaltic pump frequency are described, and the results from the data collected are detailed and commented on.

\subsubsection{Experimental Data-collection}

\begin{figure*}
\begin{minipage}[b]{0.25\linewidth} 
\centering
\includegraphics[width=1.15\linewidth,trim=0cm 1.1cm 0cm 0cm,clip]{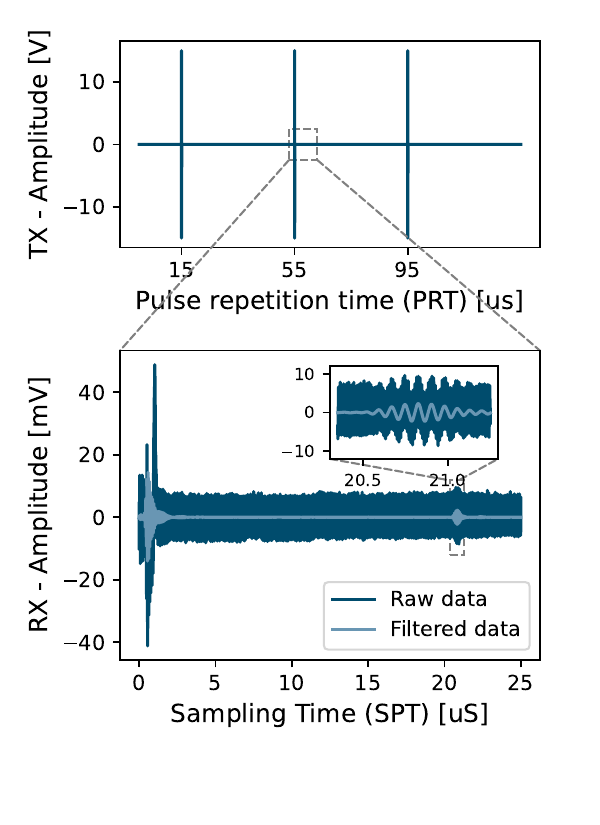}
\caption*{(a)}
\end{minipage}
\begin{minipage}[b]{0.75\linewidth} 
\centering
\begin{minipage}[b]{0.32\linewidth} 
\vspace{-0.1cm}
\includegraphics[width=1.05\linewidth]{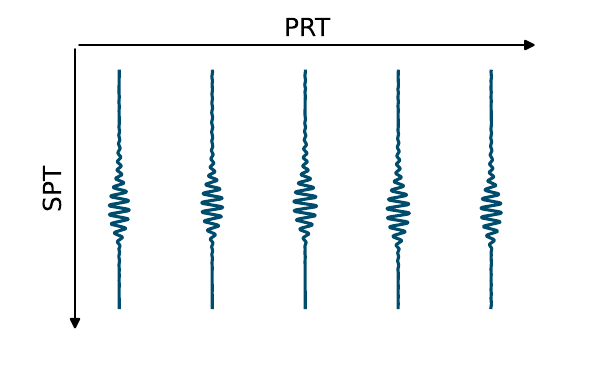}
\caption*{(b)}
\end{minipage}
\begin{minipage}[b]{0.32\linewidth} 
\includegraphics[width=1.02\linewidth]{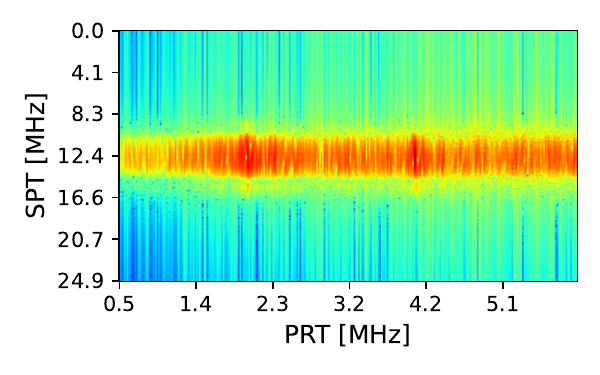}
\caption*{(d)}
\end{minipage}
\begin{minipage}[b]{0.32\linewidth} 
\includegraphics[width=1.02\linewidth,trim=1cm 0.5cm 0cm 0cm,clip]{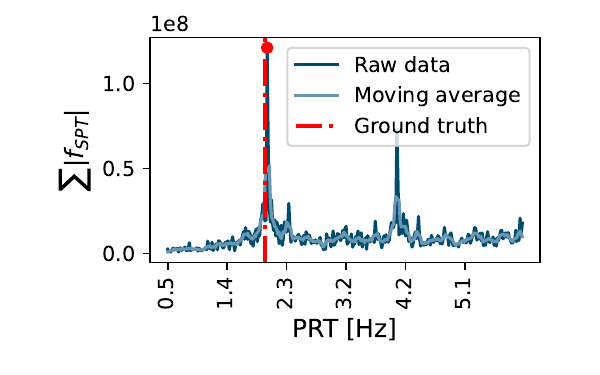}
\caption*{(f)}
\end{minipage}
\begin{minipage}[b]{0.32\linewidth} 
\vspace{0.2cm}
\includegraphics[width=1.05\linewidth]{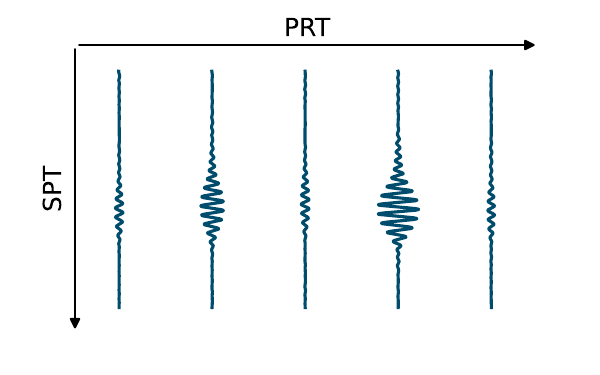}
\vspace{-0.4cm}
\caption*{(c)}
\end{minipage}
\begin{minipage}[b]{0.32\linewidth} 
\includegraphics[width=1.02\linewidth]{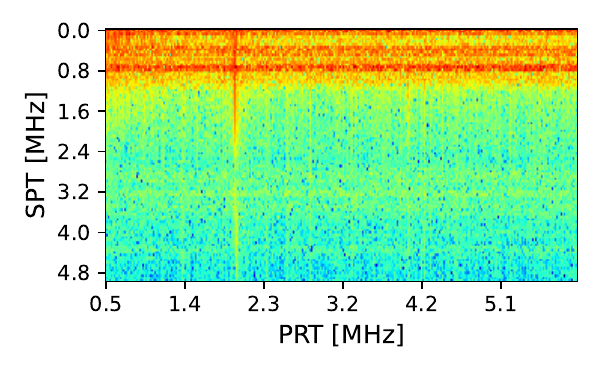}
\caption*{(e)}
\end{minipage}
\begin{minipage}[b]{0.32\linewidth} 
\includegraphics[width=1.02\linewidth,trim=1cm 0.5cm 0cm 0cm,clip]{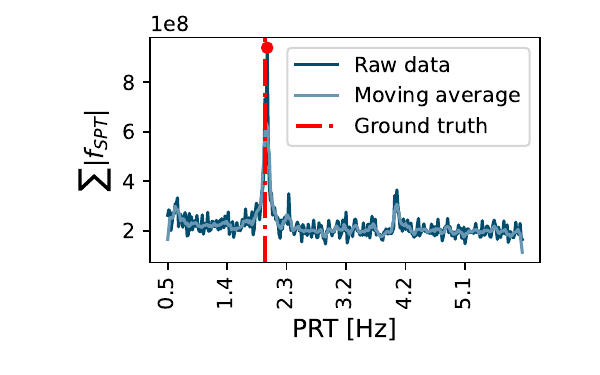}
\caption*{(g)}
\end{minipage}
\end{minipage}
\caption{Signal processing pipeline. (a) shows the raw signal as collected by the ADC, with a zoom on the US TX and RX. (b) shows different pulses of the US RX signal, in (c) it is differentiated. (d) shows a 2D FFT on the differentiated signal, (e) on the enveloped signal. (f) shows the frequency accumulation for the non-enveloped signal, (g) for the enveloped signal.}
\label{fig:sig_proc}
\vspace{-0.5cm}
\end{figure*}

For data collection, we placed our P(VDF-TrFE) transducer on top of the flow phantom. The active transducer surface was centered on the conduit of the phantom. Transmission pulses were selected as a 5-cycle \SI{12.5}{\mega\hertz} square-wave signal with a 30Vpp amplitude, and is shown on a coarse time axes in \autoref{fig:sig_proc}(a). The same figure shows also the receiving signal on the bottom, where the expanded segment highlights the reflected signal between the impedance discontinuity between the agar-agar and air.

\subsubsection{Signal acquisition}
In this work, we targeted a benchtop setup, as an explorative step towards system integration. In our experiment, the transmit pulse, the echo, and the hardware-extracted envelope of the echo have been acquired with a PCi ADC card ADQ412 (Teledyne, Sweden). An overview of the involved signals is given in \autoref{fig:sig_proc}(a).

\subsubsection{Signal processing}
To extract peristaltic pump frequencies, we adapted the method proposed in \cite{c:vostrikov2023}.
\begin{itemize}
    \item \textbf{Filtering:} the raw signal is bandpass filtered with a third order Butterworth filter with a lower bound of \SI{10}{\mega\hertz} and an upper bound of \SI{15}{\mega\hertz}.
    \item \textbf{M-Mode image:} the acquired data are put in a matrix, with each collected ultrasound shot occupying one column. This aligns the Sampling Time (SPT) along the vertical axes and the Pulse Repetition Time (PRT) along the horizontal axes, as shown in \autoref{fig:sig_proc}(b).
    \item \textbf{Differentiation:} in order to enhance the signal, the velocity of the signal is computed by differentiating along the Pulse Repetition Time axes. The output of this step can be seen in \autoref{fig:sig_proc}(c)
    \item \textbf{2D FFT:} to extract the target frequency two Fast Fourier Transforms are performed on the differentiated matrix form of the signal. 
    \autoref{fig:sig_proc}(d) and (e) report the absolute value of said FFT for the raw and enveloped echo respectively.
    \item \textbf{Frequency accumulation}: this step reduces the 2D FFT dimensionality to 1D and helps to highlight eventual time-periodic behavior by summing the absolute value of the 2D FFT along the Pulse Repetition axis. The resulting plot can be seen in \autoref{fig:sig_proc}(f) and (g) for the raw and enveloped echo respectively.
    \item \textbf{Moving average and peak finding:} the final step in signal processing is to smoothen the accumulated FFT and find the peak corresponding to our wanted frequency. In \autoref{fig:sig_proc}(d) and (g) a moving average is applied on the accumulated FFT before a peak finding algorithm is run, the extracted peak is highlighted in red.
\end{itemize}


\section{Results and Discussion}\label{sec:results}
\subsection{Flow Measurements}
To validate our setup we collected 60 seconds of ultrasound data, with a pulse repetition frequency of \SI{25}{\hertz}. We tested three different peristaltic pump frequencies, namely \SI{1}{\hertz}, \SI{1.5}{\hertz} and \SI{2}{\hertz}, which replicate common heart rate values of 60bpm, 90bpm and 120bpm.

Starting the analysis from the raw signal, shown in \autoref{fig:sig_proc}(a), it is clear how the system can be heavily duty-cycled: our application requires only \SI{25}{\micro\second} of active acquisition time over a pulse repetition period of \SI{40}{\milli\second}. 
Moving on to the M-Mode plot, \autoref{fig:sig_proc}(b), we can see how the differentiation manages to amplify the time-dependency along the PRT axes: from a slightly observable time dependency to a quite explicit time dependency in \autoref{fig:sig_proc}(c).
Focusing now on \autoref{fig:sig_proc}(d), we can observe how most of the frequency components of the signal along the SPT axes are centered, as expected, around the transducer's pulsing frequency: \SI{12.5}{\mega\hertz}. In the same plot, along the PRT axes higher values of the absolute values of the FFT are detectable at around \SI{2}{\hertz} and \SI{4}{\hertz}, with the former being the peristaltic pump frequency selected for this experiment. The same plot is also proposed in \autoref{fig:sig_proc}(e) for the enveloped signal. Focusing on the SPT axes, it is clear how the bandwidth reduction has been successful, given that the majority of the signal lies within \SI{1}{\mega\hertz}, and the pulse repetition frequencies are clearly spanning up to about \SI{2}{\mega\hertz}.
Accumulating across the PRT axes highlights the frequency components of the peristaltic pump even more, as seen in \autoref{fig:sig_proc}(f) and (g). From these plots, it is clear how the envelop detection manages to successfully reconstruct the frequency of the peristaltic pump. An additional moving window filter further helps to manage additional noise that could leak into the frequency analysis. The peaks displayed in \autoref{fig:sig_proc}(f) and (g), which corresponds to the detected peristaltic pump frequency, are clearly distinguishable and with an excellent signal-to-noise ratio. The experiment in \autoref{fig:sig_proc} had the peristaltic pump frequency set at \SI{2}{\hertz}, which is well reconstructed by the frequency accumulation. Moreover, the first harmonic of the signal at \SI{4}{Hz} is also visible.
Table \ref{tab:results_statistics} summarises the experimental runs on the three different peristaltic pump frequencies. The signal processing is introduced in \autoref{sec:methods}. A is applied on both the raw echo and the hardware-enveloped echo, and the results from the extracted peaks have been reported for each of the three selected frequencies. The set frequencies on the peristaltic pump were reconstructed, both for the raw signal and the enveloped one, within an error of \SI{0.05}{\hertz} (equivalent to \SI{3}{bpm}), which lies within errors reported by literature\cite{risso2021robust} for similar tasks. Moreover, we can observe how the envelope detection managed to reconstruct the pump frequency as good as, or even better, than the non-enveloped signal.

 We have observed that enveloping the signal can reduce the bandwidth of over 6x: in our analysis, most of the enveloped signal was in fact within the \SI{2}{\mega\hertz} mark (\autoref{fig:sig_proc}(e), while the raw echo had most of the signal above \SI{10}{\mega\hertz} (\autoref{fig:sig_proc}(d)). Reducing the bandwidth allows using MCU’s integrated analog front ends, and allows MCUs to process the amount of data acquired. This represents a lower power, cheaper, and more scalable solution than using FPGA to acquire and process ultrasound signals. The power consumption estimation breakdown for the proposed system is reported in Table \ref{tab:pow_estimation}: all the power involved are in the mW-range.
 
\vspace{-0.1cm}

\section{Conclusion}\label{sec:conclusion}

This paper proposes a methodology to lower ultrasound signal acquisition and processing requirements by enveloping the received echo. A P(VDF-TrFE)-based paper-thin flexible transducer has been used to acquire an ultrasound signal, pushing ultrasound application towards on-body continuous monitoring. A drawback of such a transducer is the higher operating frequency, which we addressed with a hardware envelope-detection circuit. An agar-agar phantom powered by a peristaltic pump has been constructed to simulate a hemodynamic application. 
Data collected on the phantom demonstrate little to no degradation in the recovery of the peristaltic pump frequency, while the required bandwidth has been decreased by over 6x. The reduced bandwidth of the hardware pre-processed signal can be then acquired and processed by common-off-the-shelf microcontrollers, drastically lowering the cost and complexity of ultrasound systems, and enabling their deployment in wearable devices.

\FloatBarrier

\section*{ACKNOWLEDGMENT}
STMicroelectronics for the provision of hardware and support. This work was supported by the ETH Research Grant ETH-C-01-21-2 (Project ListenToLight)

\bibliographystyle{IEEEtran} 
\bibliography{references.bib}

\begin{thebibliography}{10}
\providecommand{\url}[1]{#1}
\csname url@samestyle\endcsname
\providecommand{\newblock}{\relax}
\providecommand{\bibinfo}[2]{#2}
\providecommand{\BIBentrySTDinterwordspacing}{\spaceskip=0pt\relax}
\providecommand{\BIBentryALTinterwordstretchfactor}{4}
\providecommand{\BIBentryALTinterwordspacing}{\spaceskip=\fontdimen2\font plus
\BIBentryALTinterwordstretchfactor\fontdimen3\font minus \fontdimen4\font\relax}
\providecommand{\BIBforeignlanguage}[2]{{%
\expandafter\ifx\csname l@#1\endcsname\relax
\typeout{** WARNING: IEEEtran.bst: No hyphenation pattern has been}%
\typeout{** loaded for the language `#1'. Using the pattern for}%
\typeout{** the default language instead.}%
\else
\language=\csname l@#1\endcsname
\fi
#2}}
\providecommand{\BIBdecl}{\relax}
\BIBdecl

\bibitem{j:lin2023}
\BIBentryALTinterwordspacing
M.~Lin \emph{et~al.}, ``\BIBforeignlanguage{en}{A fully integrated wearable ultrasound system to monitor deep tissues in moving subjects},'' \emph{\BIBforeignlanguage{en}{Nature Biotechnology}}, pp. 1--10, May 2023, publisher: Nature Publishing Group. [Online]. Available: \url{https://www.nature.com/articles/s41587-023-01800-0}
\BIBentrySTDinterwordspacing

\bibitem{c:frey2022}
S.~Frey \emph{et~al.}, ``\BIBforeignlanguage{English}{{WULPUS}: a {Wearable} {Ultra} {Low}-{Power} {Ultrasound} probe for multi-day monitoring of carotid artery and muscle activity},'' in \emph{\BIBforeignlanguage{English}{2022 {IEEE} {International} {Ultrasonics} {Symposium} ({IUS})}}.\hskip 1em plus 0.5em minus 0.4em\relax Italy: IEEE, 2022, p.~4.

\bibitem{c:brausch2019}
\BIBentryALTinterwordspacing
L.~Brausch, H.~Hewener, and P.~Lukowicz, ``Towards a wearable low-cost ultrasound device for classification of muscle activity and muscle fatigue,'' in \emph{Proceedings of the 2019 {ACM} {International} {Symposium} on {Wearable} {Computers}}, ser. {ISWC} '19.\hskip 1em plus 0.5em minus 0.4em\relax New York, NY, USA: Association for Computing Machinery, Sep. 2019, pp. 20--22. [Online]. Available: \url{https://dl.acm.org/doi/10.1145/3341163.3347749}
\BIBentrySTDinterwordspacing

\bibitem{j:hu2023}
\BIBentryALTinterwordspacing
H.~Hu \emph{et~al.}, ``\BIBforeignlanguage{en}{A wearable cardiac ultrasound imager},'' \emph{\BIBforeignlanguage{en}{Nature}}, vol. 613, no. 7945, pp. 667--675, Jan. 2023, number: 7945 Publisher: Nature Publishing Group. [Online]. Available: \url{https://www.nature.com/articles/s41586-022-05498-z}
\BIBentrySTDinterwordspacing

\bibitem{j:wang2022}
\BIBentryALTinterwordspacing
C.~Wang \emph{et~al.}, ``Bioadhesive ultrasound for long-term continuous imaging of diverse organs,'' \emph{Science}, vol. 377, no. 6605, pp. 517--523, Jul. 2022, publisher: American Association for the Advancement of Science. [Online]. Available: \url{https://www.science.org/doi/10.1126/science.abo2542}
\BIBentrySTDinterwordspacing

\bibitem{j:yin2022}
Z.~Yin \emph{et~al.}, ``\BIBforeignlanguage{eng}{A {Wearable} {Ultrasound} {Interface} for {Prosthetic} {Hand} {Control}},'' \emph{\BIBforeignlanguage{eng}{IEEE journal of biomedical and health informatics}}, vol.~26, no.~11, pp. 5384--5393, Nov. 2022.

\bibitem{ax:grandi2023}
\BIBentryALTinterwordspacing
B.~Grandi~Sgambato \emph{et~al.}, ``\BIBforeignlanguage{en}{High {Performance} {Wearable} {Ultrasound} as a {Human}-{Machine} {Interface} for wrist and hand kinematic tracking},'' Jun. 2023. [Online]. Available: \url{https://www.techrxiv.org/articles/preprint/High_Performance_Wearable_Ultrasound_as_a_Human_Machine_Interface_for_wrist_and_hand_kinematic_tracking/22127435/2}
\BIBentrySTDinterwordspacing

\bibitem{c:vostrikov2023}
S.~Vostrikov, L.~Benini, and A.~Cossettini, ``Complete {Cardiorespiratory} {Monitoring} via {Wearable} {Ultra}-{Low} {Power} {Ultrasound},'' in \emph{2023 {IEEE} {International} {Ultrasonics} {Symposium} ({IUS})}.\hskip 1em plus 0.5em minus 0.4em\relax Montreal: IEEE, Sep. 2023.

\bibitem{c:leitner2022}
C.~Leitner \emph{et~al.}, ``Properties of a {Fully} {Printed} {Ultrasound} {Transducer} on {Flexible} {Substrate},'' in \emph{2022 {IEEE} {International} {Ultrasonics} {Symposium} ({IUS})}, Oct. 2022, pp. 1--3, iSSN: 1948-5727.

\bibitem{pvdf_advantages}
J.~Yan \emph{et~al.}, ``A lightweight ultrasound probe for wearable human–machine interfaces,'' \emph{IEEE Sensors Journal}, vol.~19, no.~14, pp. 5895--5903, 2019.

\bibitem{j:wagle2013}
\BIBentryALTinterwordspacing
S.~Wagle, A.~Decharat, P.~Bodö, and F.~Melandsø, ``Ultrasonic properties of all-printed piezoelectric polymer transducers,'' \emph{Applied Physics Letters}, vol. 103, no.~26, p. 262902, Dec. 2013. [Online]. Available: \url{https://aip.scitation.org/doi/full/10.1063/1.4857795}
\BIBentrySTDinterwordspacing

\bibitem{j:brown2000}
L.~Brown, ``Design considerations for piezoelectric polymer ultrasound transducers,'' \emph{IEEE Transactions on Ultrasonics, Ferroelectrics, and Frequency Control}, vol.~47, no.~6, pp. 1377--1396, Nov. 2000.

\bibitem{c:leitner2023}
C.~Leitner, K.~Keller, F.~Greco, and B.~Luca, ``Substrates as {Acoustic} {Modulators} in {Fully} {Printed} {Polymer} {Ultrasound} {Transducers}.'' in \emph{Proceedings of the 2023 {IEEE} {International} {Ultrasonics} {Symposium} ({IUS})}.\hskip 1em plus 0.5em minus 0.4em\relax Montreal: IEEE, Sep. 2023.

\bibitem{ultrasound_fpga}
\BIBentryALTinterwordspacing
W.~Yong \emph{et~al.}, ``Developing medical ultrasound imaging application across gpu, fpga, and cpu using oneapi,'' in \emph{International Workshop on OpenCL}, ser. IWOCL'21.\hskip 1em plus 0.5em minus 0.4em\relax New York, NY, USA: Association for Computing Machinery, 2021. [Online]. Available: \url{https://doi.org/10.1145/3456669.3456680}
\BIBentrySTDinterwordspacing

\bibitem{EARLE201618}
\BIBentryALTinterwordspacing
M.~Earle, G.~D. Portu, and E.~DeVos, ``Agar ultrasound phantoms for low-cost training without refrigeration,'' \emph{African Journal of Emergency Medicine}, vol.~6, no.~1, pp. 18--23, 2016. [Online]. Available: \url{https://www.sciencedirect.com/science/article/pii/S2211419X15001111}
\BIBentrySTDinterwordspacing

\bibitem{risso2021robust}
M.~Risso \emph{et~al.}, ``Robust and energy-efficient ppg-based heart-rate monitoring,'' in \emph{2021 IEEE International Symposium on Circuits and Systems (ISCAS)}.\hskip 1em plus 0.5em minus 0.4em\relax IEEE, 2021, pp. 1--5.

\end{thebibliography}

\end{document}